\documentclass[aps,prd,nofootinbib,superscriptaddress,twocolumn,floatfix]{revtex4}

\usepackage{times} \usepackage{pdfsync}

\newcommand{\flux}{\,\unit{cts/sec/cm^2}}

\usepackage{amssymb,amsfonts,amsmath,paralist,xspace} %

\usepackage{longtable}%
\usepackage[normalem]{ulem} %
\usepackage{relsize,color,xcolor} %
\usepackage{booktabs} 

\usepackage{units} 

\usepackage{grffile} %
\usepackage{graphicx} %
\graphicspath{{./figures/}} 

\IfFileExists{srcltx.sty}{\usepackage[active]{srcltx}}

\usepackage{etoolbox} %
\preto\tabular{\setcounter{magicrownumbers}{0}}%
\newcounter{magicrownumbers}%
\newcommand\rownumber{\stepcounter{magicrownumbers}\arabic{magicrownumbers}}%

 \newcommand{\be}
{\begin{equation}} \newcommand{\ee} {\end{equation}}   
\newcommand{\eq}[1]{\begin{equation} #1 \end{equation}}
\def\be{\begin{eqnarray}} \def\ee{\end{eqnarray}} 
 \renewcommand{\S}{\mathcal{S}} %
\newcommand{\dm}{{\textsc{dm}}} 

\renewcommand{\S}{\mathcal{S}} 

\newcommand{\nfw}{\mathrm{NFW}} 
\newcommand{\iso}{\mathrm{ISO}} 
\begin{document}


\title{Checking the dark matter origin of 3.53~keV line with the Milky Way
  center}

\author{A.~Boyarsky$^{1}$, J.~Franse$^{1,2}$, D.~Iakubovskyi$^{3}$, and O.~Ruchayskiy$^{4}$ \\
  $^1${\small Instituut-Lorentz for Theoretical Physics, Universiteit Leiden,
    Niels Bohrweg 2, Leiden, The Netherlands}\\
  $^2${\small Leiden Observatory, Leiden University, Niels Bohrweg 2, Leiden, The Netherlands}\\
  $^3${\small Bogolyubov Institute of Theoretical Physics, Metrologichna
    Str. 14-b, 03680, Kyiv, Ukraine}\\
  $^4${\small Ecole Polytechnique F\'ed\'erale de Lausanne, FSB/ITP/LPPC, BSP, CH-1015, Lausanne, Switzerland}\\
} \date{\today}

\begin{abstract}
  We detect a line at $3.539\pm 0.011$~keV in the deep exposure dataset of the
  Galactic Center region, observed with the XMM-Newton.  
  The dark matter interpretation of the signal observed
  in 
  the Perseus galaxy cluster, the Andromeda galaxy~\cite{Boyarsky:14a} and in
  the stacked spectra of galaxy clusters~\cite{Bulbul:14}, together with
  non-observation of the line in blank sky data, put both lower and upper
  limits on the possible intensity of the line in the Galactic Center
  data. Our result is consistent with these constraints for a class of Milky
  Way mass models, presented previously by observers, and would correspond to radiative decay dark matter
  lifetime $\tau_\dm \sim 6-8\times 10^{27}$~sec. Although it is hard to
  exclude an astrophysical origin of this line based the Galactic Center data
  alone, this is an important consistency check of the hypothesis that
  encourages to check it with more observational data that are expected by the
  end of 2015.
\end{abstract}

\maketitle


Recently, two independent groups~\cite{Bulbul:14,Boyarsky:14a} reported a
detection of an unidentified X-ray line at energy $3.53$~keV in the
long-exposure X-ray observations of a number of dark matter-dominated objects.
The authors of~\cite{Bulbul:14} have observed this line in a stacked XMM spectrum of
73 galaxy clusters spanning a redshift range $0.01-0.35$ and separately in
subsamples of nearby and remote clusters.  Ref.~\cite{Boyarsky:14a} have found
this line in the outskirts of the Perseus cluster and in the central $14'$ of
the Andromeda galaxy. The global significance of detection of the same line in
the datasets of Ref.~\cite{Boyarsky:14a} is $4.3\sigma$ (taking into account
the trial factors); the signal in~\cite{Bulbul:14} has significance above
$4\sigma$ based on completely independent data.

The position of the line is correctly redshifted between galaxy
clusters~\cite{Bulbul:14} and between the Perseus cluster and the Andromeda
galaxy~\cite{Boyarsky:14a}.  In a very long exposure blank sky observation
(15.7 Msec of cleaned data) the feature is absent~\cite{Boyarsky:14a}.  This
makes it unlikely that an instrumental effect is at the origin of this feature (e.g.\ an unmodeled
wiggle in the effective area).

To identify this spectral feature with an atomic line in galaxy clusters, one
should assume a strongly super-solar abundance of potassium or some anomalous
argon transition~\cite{Bulbul:14}. Moreover, according to the results
of~\cite{Boyarsky:14a} this should be true not only in the center of the
Perseus cluster considered in~\cite{Bulbul:14}, but also \textit{(i)} in its
outer parts up to at least 1/2 of its virial radius and \textit{(ii)} in the
Andromeda galaxy.

This result triggered significant interest as it seems consistent with a
long-sought-for signal from dark matter
decay~\cite{Ishida:14a,Higaki:14,Jaeckel:14,Czerny:14,Lee:14a,Abazajian:14,
  Krall:14,ElAisati:14,Frandsen:14,Hamaguchi:14,Kong:14,Baek:14a,Nakayama:14a,Choi:14,
  Shuve:14,Kolda:14,Allahverdi:14,Dias:14,Bomark:14,PeiLiew:14,Nakayama:14b,Kang:14,Okada:14a,
  Demidov:14,Queiroz:14,Babu:14,Prasad:14,Rosner:14,Barry:14,
  Robinson:14,Kubo:14,Drewes:14,Baek:14b,Nakayama:14c,Bateman:14,Chakraborty:14,Lattanzi:14,
  Kawasaki:14,Chen:14,Ishida:14b,Abada:14a,Abada:14b,Baer:14,Ringwald:14,
  Dutta:14}, 
annihilation~\cite{Frandsen:14,Dudas:14,Baek:14b}, 
de-excitation~\cite{Finkbeiner:14,Cline:14a,Frandsen:14,Chiang:14,Geng:14,Okada:14b,Lee:14b,Falkowski:14,
Cline:14b,Boddy:14,Mambrini:15} or
conversion in the magnetic field~\cite{Conlon:14b,Cicoli:14,Conlon:14a}. Many
particle physics models that predict such properties for the dark matter
particle, have been put forward, including sterile neutrino, axion, axino,
gravitino and many others, see for reviews
e.g.\ ~\cite{Iakubovskyi:14,Bateman:14} and references therein. If the interaction of
dark matter particles is weak enough (e.g.\ much weaker than that of the
Standard Model neutrino), they need not to be stable as their lifetime can
exceed the age of the Universe. Nevertheless huge amounts of dark matter
particles can make the signal strong enough to be detectable even from such
rare decays.

The omni-presence of dark matter in galaxies and galaxy clusters opens the way
to check the decaying dark matter hypothesis~\cite{Boyarsky:10b}.  The
decaying dark matter signal is proportional to the \emph{column density}
$\S_\dm=\int \rho_\dm d\ell$ -- the integral along the line of sight of the DM
density distribution (unlike the case of annihilating dark matter, where the
signal is proportional to $\int \rho_\dm^2 d\ell$). As long as the angular
size of an object is larger than the field-of-view, the distance to the object
drops out which means that distant objects can give fluxes comparable to those of nearby
ones~\cite{Boyarsky:09c,Boyarsky:09b}.
It also does not decrease with the distance from the centres of objects as
fast as e.g. in the case of annihilating DM where the expected signal is
concentrated towards the centers of DM-dominated objects.  This in principle
allows one to check the dark matter origin of a signal by comparison between
objects and/or by studying the angular dependence of the signal within one object,
rather than trying to exclude all possible astrophysical explanations for each
target~\cite{Boyarsky:06c,denHerder:09,Abazajian:09a,Boyarsky:12c}.

 \begin{figure*}[tp!]
  \centering
  \includegraphics[width=0.45\textwidth]{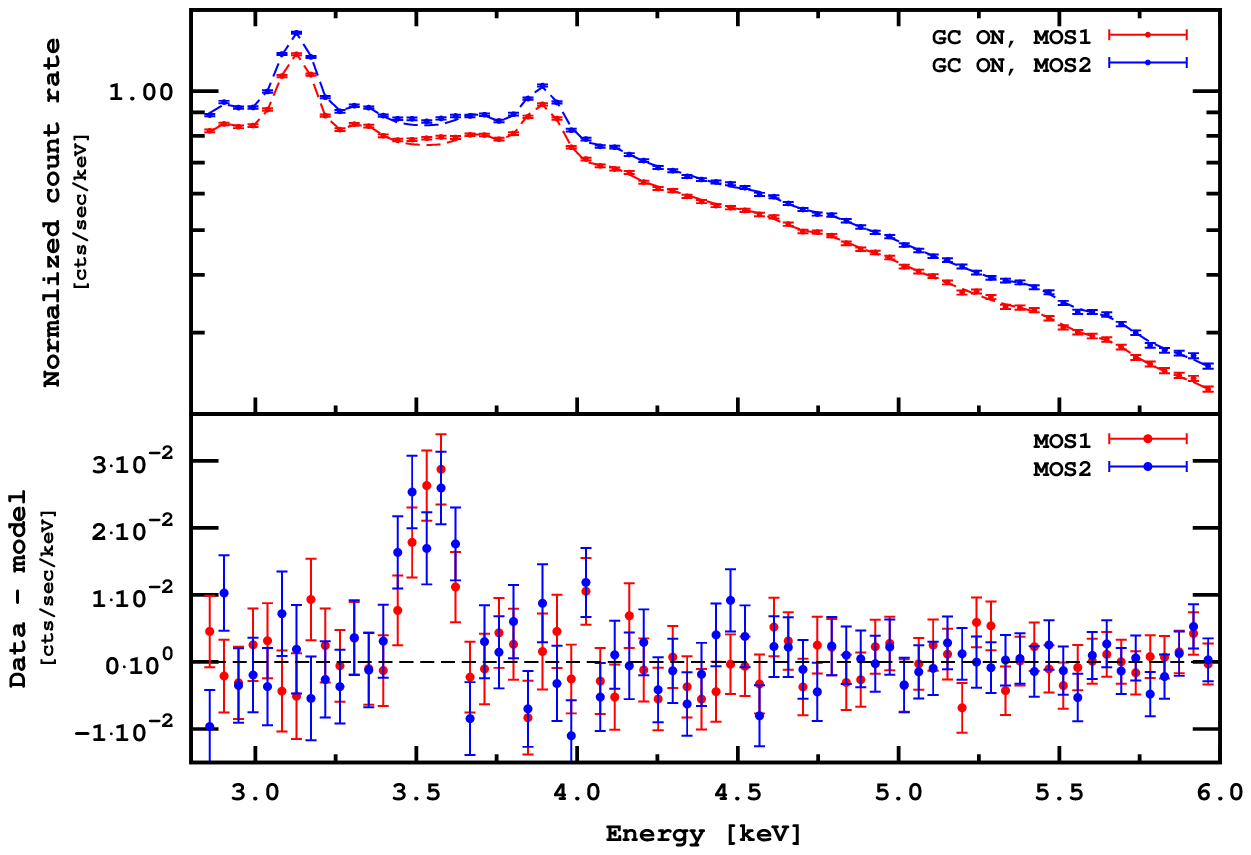}~%
  \includegraphics[width=0.45\textwidth]{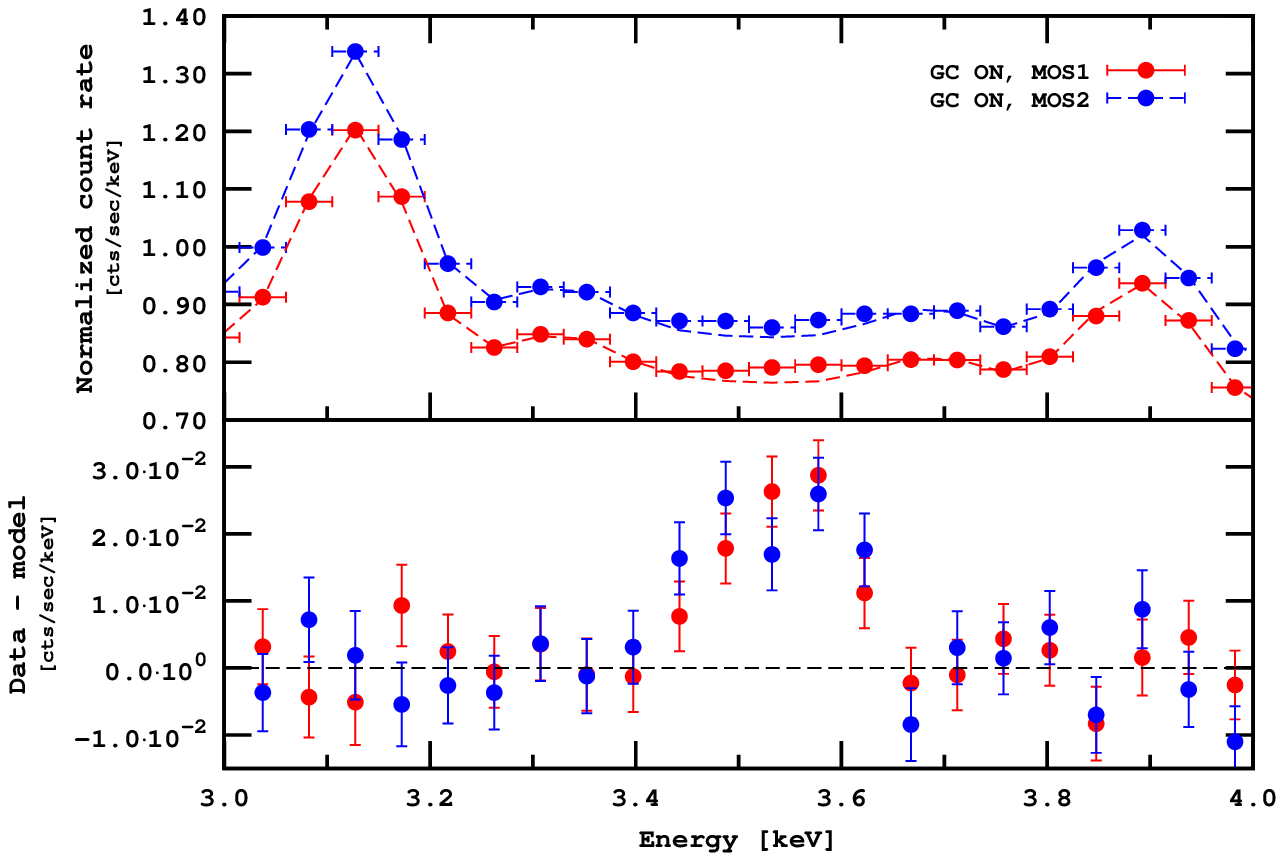}
  \caption{\textit{Left:} Folded count rate for MOS1 (lower curve, red) and
    MOS2 (upper curve, blue) and residuals (bottom) when the line at
    $3.54$~keV is \emph{not added}. The difference between the cameras is due to detector gaps and bad pixels. 
    \textit{Right}: Zoom at the range 3.0--4.0
    keV.}
  \label{fig:gc-residuals}
\end{figure*}

Clearly, after years of systematic searches for this
signal
(Ref.~\cite{Abazajian:01b,Dolgov:00,Boyarsky:05,Boyarsky:06b,Boyarsky:06c,Riemer-Sorensen:06a,Watson:06,
  Riemer-Sorensen:06b,Boyarsky:06d,Abazajian:06,Boyarsky:06e,Boyarsky:06f,Yuksel:07,Boyarsky:07a,
  Boyarsky:07b,Loewenstein:08,Riemer-Sorensen:09,Loewenstein:09,Boyarsky:10a,Mirabal:10a,Mirabal:10b,
  Prokhorov:10,Borriello:11,Watson:11,Loewenstein:12,Kusenko:12,Horiuchi:13};
see Fig.~4 in~\cite{Boyarsky:14a}) any
candidate line can be detected only at the edge of the possible sensitivity of
the method. Therefore, to cross-check the signal one needs 
long-exposure data. Moreover, even a factor 2 uncertainty in the expected signal (which is impossible to avoid) 
can result in the
necessity to have significantly more statistics than in the initial data set
in which the candidate signal was found.

So far the DM interpretation of the signal of ~\cite{Bulbul:14} and
~\cite{Boyarsky:14a} is consistent with the data: it has the correct scaling
between the Perseus cluster, Andromeda and the upper bound from the
non-detection in the blank sky data~\cite{Boyarsky:14a}, and between different
subsamples of clusters~\cite{Bulbul:14}. The mass and lifetime of the dark matter
particle that is implied by the DM interpretation of the results of~\cite{Boyarsky:14a}, is
consistent with the results of~\cite{Bulbul:14}. The signal has radial
surface brightness profiles in the Perseus cluster and Andromeda~\cite{Boyarsky:14a} that are consistent 
with a dark matter
distribution. Although the significance of this result is not sufficient to
confirm the hypothesis, they can be considered as successful sanity
checks. More results are clearly needed to perform a convincing checking
program as described above.

A classical target for DM searches is the centre of our Galaxy. Due to its proximity
it is possible to concentrate on the very central part and therefore, even for
decaying DM, one can expect a significant gain in the signal if the DM
distribution in the Milky Way happens to be steeper than a cored profile. The
Galactic Center (GC) region has been extensively studied by the XMM and
several mega-seconds of raw exposure exist. On the other hand, the GC region
has strong X-ray emission as many complicated processes occur
there~\cite{Koyama:89,Koyama:96,Kaneda:97,Muno:04,
  Koyama:06,Muno:06,Revnivtsev:09,Ponti:10,Uchiyama:12}. In particular, the
X-ray emitting gas may contain several thermal components with different
temperatures; it may be more difficult to constrain the abundances of
potassium and argon reliably than in the case of intercluster medium. Therefore the GC
data alone would hardly provide a convincing detection of the DM signal, as even
a relatively strong candidate line could be explained by astrophysical
processes.  In this paper we pose a different question: \emph{Are the
  observations of the Galactic Center consistent with the dark matter
  interpretation of the $3.53$~keV line of~\cite{Bulbul:14,Boyarsky:14a}?}

The DM interpretation of the $3.53$~keV line in M31 and Perseus
provides a prediction of the \emph{minimal} expected flux from the GC. On the other hand, the
non-detection of any signal in the off-center observations of the Milky Way
halo (the blank sky dataset of~\cite{Boyarsky:14a}) provides the prediction of the \emph{maximal} 
possible flux in the GC, given observational constraints on
the DM distribution in the Galaxy. Therefore, even with all the uncertainties
on the DM content of the involved objects, the expected signal from the GC is
bounded from both sides and provides a non-trivial check for the DM
interpretation of the $3.53$~keV line.

We use XMM-Newton observations of the central $14'$ of the Galactic Center
region with a total cleaned exposure of 1.4~Msec.
We find that the spectrum has a $\sim 5.7\sigma$ line-like excess at the expected
energy. The simultaneous fitting of the GC, Perseus and M31 provides a $\sim
6.7\sigma$ significant signal at the same position, with the detected fluxes
being consistent with the DM interpretation. The fluxes are also consistent
with the non-observation of the signal in the blank-sky and M31 off-center
datasets, if one assumes a steeper-than-cored DM profile (for example, the NFW profile of
Ref.~\cite{Weber:09}).

Below we summarize the details of our data analysis and discuss the results.

\textbf{Data reduction.}  We use all archival data of the Galactic Center
obtained by the EPIC MOS cameras~\cite{Turner:00} with Sgr A* less than $0.5'$
from the telescope axis (see SOM, Table~I).  The
data are reduced by the standard \tt SAS\rm\footnote{v.13.5.0 \tt
  http://xmm.esa.int/sas\rm} pipeline, including screening for the
time-variable soft proton flares by \texttt{espfilt}. We removed the
observations taken during the period MJD 54000--54500 due to strong flaring activity
of Sgr A* (see SOM, Fig.~1). The
data reduction and preparation of the final spectra are similar
to~\cite{Boyarsky:14a}.  For each reduced observation we select a circle of
radius $14'$ around Sgr A* and combine these spectra using the
\texttt{FTOOLS}~\cite{ftools} procedure \texttt{addspec}.

\textbf{Spectral modeling.}  To account for the cosmic-ray induced
instrumental background we have subtracted the latest closed filter datasets
(exposure: 1.30~Msec for MOS1 and 1.34~Msec for MOS2)~\cite{closed-filter}.
The rescaling of the closed filter data has been performed such that the
flux at energies $E>10$~keV reduces to zero (see~\cite{Nevalainen:05} for details).  We model
the resulting physical spectrum in the energy range 2.8--6.0 keV.  The X-ray
emission from the inner part of the Galactic Center contains both thermal and
non-thermal components~\cite{Kaneda:97,Muno:04}. Therefore, we chose to model
the spectrum with a thermal plasma model (\tt vapec\rm) and a non-thermal
\texttt{powerlaw} component modified by the \texttt{phabs} model to account for
the Galactic absorption.\footnote{The \texttt{Xspec}~\cite{Arnaud:96} v.12.8.0
  is used for the spectral analysis.}  We set the abundances of all elements -- except for
Fe -- to zero but model the known astrophysical lines with
\texttt{gaussians}~\cite{Bulbul:14,Boyarsky:14a,Riemer-Sorensen:14}. We
selected the $\geq 2\sigma$ lines from the set of astrophysical lines
of~\cite{Uchiyama:12,Bulbul:14}\footnote{Unlike~\cite{Bulbul:14} we do not
  include K~XVIII lines at 3.47 and 3.51~keV to our model. {See the discussion
    below}}. The intensities of the lines are allowed to vary, as are the
central energies to account for uncertainties in detector gain and limited
spectral resolution. We keep the same position of the lines between the two
cameras.

The spectrum is binned to 45~eV to have about 4 bins per resolution element.
The fit quality for the dataset is $\chi^2 = 108/100$~d.o.f.  The resulting
values for the main continuum components -- the folded \texttt{powerlaw} index
(for the integrated point source contribution), the temperature of the
\texttt{vapec} model ($\sim$8 keV), and the absorption column density -- agree
well with previous studies~\cite{Kaneda:97,Muno:04}.

\textbf{Results.} The resulting spectra of the inner $14'$ of the Galactic
Center show a $\sim 5.7\sigma$ line-like excess at $3.539\pm 0.011$~keV with
a flux of $(29\pm 5)\times 10^{-6}\flux$ (see Fig.~\ref{fig:gc-residuals}). It
should be stressed that these $1\sigma$ error-bars are obtained with the
\texttt{xspec} command \texttt{error} (see \textbf{Discussion} below). The position
of the excess is very close to the similar excesses recently observed in
Andromeda ($3.53 \pm 0.03$~keV) and Perseus ($3.50\pm 0.04$~keV) reported in~\cite{Boyarsky:14a}, and is less
than $2\sigma$ away from the one described in~\cite{Bulbul:14}.

We also performed combined fits of the GC dataset with those of M31 and Perseus
from~\cite{Boyarsky:14a}. As mentioned, the data reduction and modeling were
performed very similarly, so we suffice with repeating that the inner part of
M31 is covered by almost 1~Msec of cleaned MOS exposure, whereas a little over
500~ksec of clean MOS exposure was available for Perseus (see
~\cite{Boyarsky:14a} for details).

We first perform a joint fit to the Galactic Center and M31, and subsequently
to the Galactic Center, M31 and Perseus.  In both cases, we start with the
best-fit models of each individual analysis without any lines at $3.53$~keV,
and then add an additional gaussian to each model, allowing the energy to vary
while keeping the same position between the models.  The normalizations of
this line for each dataset are allowed to vary independently.  In this way,
the addition of the line to the combination of Galactic Center, M31 and
Perseus gives 4 extra degrees of freedom, which brings the joint significance
to $\sim 6.7\sigma$.

\begin{figure}[!t]
  \centering
  \includegraphics[width=.45\textwidth]{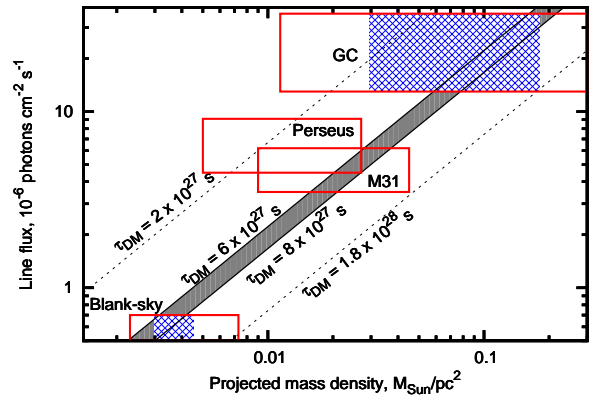}%
  \caption{The flux of the $3.53$~keV line in the spectra of the GC (this
    work), the Perseus cluster outskirts, M31, and the blank
    sky~\protect\cite{Boyarsky:14a} as a function of the DM projected mass.  Diagonal lines show the 
    expected behaviour of a
    decaying DM signal for a given DM particle lifetime.  The vertical sizes
    of the boxes are $\pm 1 \sigma$ statistical error on the line's flux -- or
    the $2\sigma$ upper bound for the blank-sky dataset.  The horizontal sizes
    of the boxes bracket the \emph{scatter} in the literature mass modeling
    (see text and Appendix A).  The Milky Way halo contribution is
    included for M31 but not for Perseus, where it would be redshifted.
    The projected mass density for the GC and the Milky Way outskirts (blank
    sky) are correlated. The blue shaded regions show a particular NFW
    profile of the Milky Way~\protect\cite{Smith:06}, its horizontal size
    indicates uncertainties in galactic disk modeling. 
    Other \emph{cuspy} profiles are consistent with these flux ratios as well (c.f.~\cite{Lovell:14}). 
    The lifetime $\tau_\dm \sim (6-8)\times 10^{27}$~sec is consistent with all datasets.}
  \label{fig:lifetime}
\end{figure}

To further investigate possible systematic errors on the line parameters we
took into account that the \texttt{gaussian} component at $3.685$~keV may
describe not a single line, but a complex of lines
(SOM, Table~II). Using the \texttt{steppar} command we scanned over
the two-dimensional grid of this \texttt{gaussian}'s intrinsic width and the
normalization of the line at $3.539$~keV. We were able to find a new best fit
with the $3.685$~keV \texttt{gaussian} width being as large as $66\pm 15$~eV.
In this new minimum our line shifts to $3.50 \pm 0.02$~keV (as some of the
photons were attributed to the $3.685$~keV \texttt{gaussian}) and has a flux of
$24\times 10^{-6}\flux$ with a $1\sigma$ confidence interval of $(13 - 36)\times
10^{-6}\flux$. The significance of the line is $\Delta \chi^2 = 9.5$
($2.6\sigma$ for 2 d.o.f.). Although the width in the new minimum seems to be
too large even for the whole complex of Ar XVII lines (see
\textbf{Discussion}), we treat this change of line parameters as the
estimate of systematic uncertainties.
To reduce these systematics one has
either to resolve or to reliably model a line complex around $3.685$~keV
instead of representing it as one wide \texttt{gaussian} component.

As was argued in~\cite{Boyarsky:14a}, an interpretation of the signal as an unmodelled wiggle in the 
effective area is not favoured because it should have produced a very significant signal in the 
blank-sky dataset as well. This is because an effect like this would produce a line-like residual 
proportional to the continuum level. In addition, the line would not be redshifted properly 
for Perseus~\cite{Boyarsky:14a} and the cluster stack from~\cite{Bulbul:14}.

\textbf{Discussion.} The intensity of DM decay signal should correlate with DM content of the probed objects. 
In order to check this we took DM distributions for Perseus, M31 and the MW from
Refs.~\cite{Widrow:05,Geehan:05,Battaglia:05,Battaglia:06,Smith:06,Weber:09,Chemin:09,Corbelli:09,
McMillan:11,Simionescu:11,Nesti:13,Xue:08}
(see Appendix for details) and plotted the line intensity vs.\ mass in the
field-of-view divided by the distance squared (\emph{projected DM density}), Fig.~\ref{fig:lifetime}. 
We see that \emph{decaying DM with a lifetime $\tau_\dm \sim 6-8\times 10^{27}$~sec would explain the signals 
from the GC, Perseus and M31 and the non-observation in the blank-sky dataset.}
A considerable spread of projected DM masses is due to scatter between the
distributions in the literature.
For the GC the estimates are based on extrapolations, as there are no measurements of the DM distribution 
within the inner few kpc. 
The correlation between the GC and blank-sky projected DM densities is necessary, since these are different 
parts of the same halo. From comparing our GC signal with the blank-sky upper limit we see that 
\emph{this requires cuspy (rather than cored) density profile of the Milky Way}.
Fig.~\ref{fig:lifetime} shows an example of a profile consisten with both the GC detection and blank-sky 
upper limit, Ref.~\cite{Smith:06}.

M31 and Milky Way are expected to have similar distributions, providing another consistency check.
Ref.~\cite{Boyarsky:14a} showed that in order to explain the signal from central $14'$ and non-observation 
from M31 outskirts, the Andromeda DM density profile should be cuspy, as predicted also for the Milky Way.
Fig.~\ref{fig:lifetime} shows that indeed large projected DM mass (i.e.\ cuspy profile) is preferred for M31. 

Finally the Perseus signal of~\cite{Boyarsky:14a} comes from the cluster outskirts where the hydrostatic 
mass~\cite{Simionescu:11} may be under-estimated~\cite{Okabe:14}. 
This would only improve the consistency between the data sets.

The comparison of expected DM signal from GC vs.\ blank-sky vs.\ Andromeda has been investigated in 
simulations~\cite{Lovell:14}, where various realisation of the galactic DM halos were considered 
and high probability of finding observed flux ratios between GC and M31 and between GC and blank-sky 
upper limit was found.

The non-detection of the signal in stacked dSphs by~\cite{Malyshev:14} rules out the central values 
of the decay lifetime from~\cite{Bulbul:14} but is consistent with~\cite{Boyarsky:14a} in case of 
large project DM mass (also preferred from comparison with other signals, Fig.~\ref{fig:lifetime}).
The signal was not detected in stacked galaxy spectra~\cite{Anderson:14}.
However, a novel method of~\cite{Anderson:14} has pronounced systematic effects (see Appendix~B 
of~\cite{Anderson:14}) and is the least sensitive exactly at energies $E\sim 3.5$~keV. 
Ref.~\cite{Iakubovskyi:14} used a stacked dataset of nearby galaxies from~\cite{Iakubovskyi:13} and 
showed that systematic effects and uncertainty in dark matter distributions~\cite{Boyarsky:09c} 
lead to the bound $\tau_\dm \gtrsim 3.5\times 10^{27}$~sec, consistent with our findings.
Other bounds on decaying dark matter in the $\sim 3.5$~keV energy range 
(see~\cite{Iakubovskyi:13,Horiuchi:13,Sekiya:15} and references therein) are also consistent with 
our detections for lifetimes that we discuss in this paper.

As mentioned in the \textbf{Results}, there is a degeneracy between the width
of the Ar~XVII complex around $3.685$~keV and the normalization of the line in
question.  If we allow the width of the Ar~XVII line to vary freely we can
decrease the significance of the line at $3.539$~keV to about $2\sigma$. However,
in this case the width of the \texttt{gaussian} at $3.685$~keV should be
$95-130$~eV, which is significantly larger than we obtain when simulating a
complex of four Ar~XVII lines.  In addition, in
this case the total flux of the line at $3.685$~keV becomes \emph{higher} than
the fluxes in the lines at $3.130$ and $3.895$ in contradiction with the
atomic data (SOM, Table~II).

Another way to decrease the significance of the line at $3.539$ is to assume
the presence of a potassium ion (K~XVIII) with a line at $3.515$~keV and a
smaller line at $3.47$~keV. If one considers the abundance of potassium as a
completely free parameter (c.f.\ \cite{Riemer-Sorensen:14,Carlson:14,Jeltema:14a}), one can find an acceptable fit of the
XMM GC data without an additional line at $3.539$~keV.
As described in Appendix B, due to the complicated internal temperature and abundance structures it 
is not possible to reliably constrain the overall potassium abundance of the GC to a degree that rules 
out the K~XVIII origin of the $3.539$~keV line \emph{in this dataset.}

However, if we are to explain the presence of this line in the spectra by the
presence of K~XVIII, we have to build a model that consistently explains the
fluxes in this line in different astronomical environments: in galaxy clusters
(in particular Perseus) at all off-center distances from the central
regions~\cite{Bulbul:14} to the cluster outskirts up to the virial
radius~\cite{Boyarsky:14a}; in the central part of M31; and in the Galactic Center.  
In addition, we need to explain that this line is not observed -- and therefore that this transition 
\emph{should not} be excited -- in the outskirts of the
Milky Way and of M31~\cite{Boyarsky:14a}. 
Such a consistent model does not look convincing. In particular, in M31 spectrum there are no strong
astrophysical lines in $3-4$~keV range~\cite{Boyarsky:14c}. The powerlaw continuum is
well determined by fitting the data over a wider range of energies (from 2 to
8 keV) and allows a clear detection of the line at $3.53\pm
0.03$~keV with $\Delta \chi^2 = 13$~\cite{Boyarsky:14a,Boyarsky:14c}, which is also the largest line-like
feature in the entire 3--4~keV range. Were this signal in M31 due to K~XVIII, there should be plenty of 
stronger emission lines present. In addition, the authors of~\cite{Bulbul:14}
conclude that strongly super-solar abundances of K~XVIII are required to explain
the observed excess of this line in their stacked cluster analysis.

\emph{In conclusion}, although it is hard to exclude completely an astrophysical
origin of the $3.539$~keV line in the GC (due to the complicated nature of this object), 
the detection of this line in this object is an essential cross-check for the DM
interpretation of the signal observed in Perseus and M31~\cite{Boyarsky:14a}
and in the stacked spectra of galaxy clusters~\cite{Bulbul:14}. A non-detection in the GC or a detection 
with high flux would have immediately ruled out this interpretation.
As it is, the GC data rather supports DM
interpretation as the line is not only observed at the same energy, but
also its flux is consistent with the expectations about the DM distributions.

To settle this question, measurements with higher spectral resolution, an
independent measurement of the relative abundances of elements in the GC
region, and analyses of additional deep exposure datasets of DM-dominated
objects are
needed~\cite{Koyama:14,Lovell:14,Figueroa-Feliciano:15,Sekiya:15,Iakubovskyi:15a,Speckhard:15}
with Astro-H~\cite{Kitayama:2014fda} or future mision, Athena~\cite{Neronov:2015kca}.

\textbf{Acknowledgments.} The work of D.~I. was supported
by part by the Swiss National Science Foundation grant SCOPE IZ7370-152581, 
the Program of Cosmic Research of the National Academy of
Sciences of Ukraine, the State Programme of Implementation of Grid
Technology in Ukraine and the grant of President of Ukraine for young scientists. 
The work of J.~F. was supported by the De Sitter program
at Leiden University with funds from NWO. This research is part of the
``Fundamentals of Science'' program at Leiden University.

\let\jnlstyle=\rm\def\jref#1{{\jnlstyle#1}}\def\aj{\jref{AJ}}
  \def\araa{\jref{ARA\&A}} \def\apj{\jref{ApJ}\ } \def\apjl{\jref{ApJ}\ }
  \def\apjs{\jref{ApJS}} \def\ao{\jref{Appl.~Opt.}} \def\apss{\jref{Ap\&SS}}
  \def\aap{\jref{A\&A}} \def\aapr{\jref{A\&A~Rev.}} \def\aaps{\jref{A\&AS}}
  \def\azh{\jref{AZh}} \def\baas{\jref{BAAS}} \def\jrasc{\jref{JRASC}}
  \def\memras{\jref{MmRAS}} \def\mnras{\jref{MNRAS}\ }
  \def\pra{\jref{Phys.~Rev.~A}\ } \def\prb{\jref{Phys.~Rev.~B}\ }
  \def\prc{\jref{Phys.~Rev.~C}\ } \def\prd{\jref{Phys.~Rev.~D}\ }
  \def\pre{\jref{Phys.~Rev.~E}} \def\prl{\jref{Phys.~Rev.~Lett.}}
  \def\pasp{\jref{PASP}} \def\pasj{\jref{PASJ}} \def\qjras{\jref{QJRAS}}
  \def\skytel{\jref{S\&T}} \def\solphys{\jref{Sol.~Phys.}}
  \def\sovast{\jref{Soviet~Ast.}} \def\ssr{\jref{Space~Sci.~Rev.}}
  \def\zap{\jref{ZAp}} \def\nat{\jref{Nature}\ } \def\iaucirc{\jref{IAU~Circ.}}
  \def\aplett{\jref{Astrophys.~Lett.}}
  \def\apspr{\jref{Astrophys.~Space~Phys.~Res.}}
  \def\bain{\jref{Bull.~Astron.~Inst.~Netherlands}}
  \def\fcp{\jref{Fund.~Cosmic~Phys.}} \def\gca{\jref{Geochim.~Cosmochim.~Acta}}
  \def\grl{\jref{Geophys.~Res.~Lett.}} \def\jcp{\jref{J.~Chem.~Phys.}}
  \def\jgr{\jref{J.~Geophys.~Res.}}
  \def\jqsrt{\jref{J.~Quant.~Spec.~Radiat.~Transf.}}
  \def\memsai{\jref{Mem.~Soc.~Astron.~Italiana}}
  \def\nphysa{\jref{Nucl.~Phys.~A}} \def\physrep{\jref{Phys.~Rep.}}
  \def\physscr{\jref{Phys.~Scr}} \def\planss{\jref{Planet.~Space~Sci.}}
  \def\procspie{\jref{Proc.~SPIE}} \let\astap=\aap \let\apjlett=\apjl
  \let\apjsupp=\apjs \let\applopt=\ao \def\jcap{\jref{JCAP}}

\appendix
\onecolumngrid

\begin{center}
  \large \textbf{Supplementary material}
\end{center}

\setcounter{magicrownumbers}{0}
\begin{table*}[!t]
\centering \relsize{-0.6}
\begin{tabular}[c]{r|c|c|c|c|l}
\hline
 & ObsID & Off-center angle & Cleaned exposure & FoV [arcmin$^2$] \\
   && arcmin & MOS1/MOS2 [ksec] & MOS1/MOS2 \\
   \hline
\rownumber & \texttt{0111350101} & 0.017 & 40.8/40.7 & 570.5/570.3 \\
\rownumber & \texttt{0111350301} & 0.017 & 7.2/6.8 & 565.8/573.4 \\
\rownumber & \texttt{0112972101} & 0.087 & 20.8/21.4 & 571.4/572.0 \\
\rownumber & \texttt{0202670501} & 0.003 & 21.4/26.5 & 564.9/573.4 \\
\rownumber & \texttt{0202670601} & 0.003 & 29.6/31.1 & 563.8/574.1 \\
\rownumber & \texttt{0202670701} & 0.003 & 76.0/80.0 & 570.4/573.3 \\
\rownumber & \texttt{0202670801} & 0.003 & 86.9/91.0 & 569.2/572.8 \\
\rownumber & \texttt{0402430301}$^{a}$ & 0.002 & 57.6/60.2 & 475.8/572.1 \\
\rownumber & \texttt{0402430401}$^{a}$ & 0.002 & 37.3/37.8 & 476.2/572.3 \\
\rownumber & \texttt{0402430701}$^{a}$ & 0.002 & 23.1/25.2 & 478.5/573.1 \\
\rownumber & \texttt{0504940201}$^{a}$ & 0.286 & 7.7/8.5 & 487.6/572.6 \\
\rownumber & \texttt{0505670101}$^{a}$ & 0.002 & 65.7/73.7 & 472.0/573.2 \\
\rownumber & \texttt{0554750401} & 0.003 & 31.6/31.5 & 483.4/574.0 \\
\rownumber & \texttt{0554750501} & 0.003 & 39.6/39.2 & 487.0/574.0 \\
\rownumber & \texttt{0554750601} & 0.003 & 35.5/36.4 & 487.0/573.3 \\
\rownumber & \texttt{0604300601} & 0.003 & 28.9/30.0 & 487.1/573.1 \\
\rownumber & \texttt{0604300701} & 0.003 & 35.1/37.1 & 487.4/572.7 \\
\rownumber & \texttt{0604300801} & 0.003 & 34.9/34.2 & 487.8/572.5 \\
\rownumber & \texttt{0604300901} & 0.003 & 21.1/20.7 & 485.1/574.0 \\
\rownumber & \texttt{0604301001} & 0.003 & 35.3/38.6 & 487.4/573.6 \\
\rownumber & \texttt{0658600101} & 0.078 & 46.5/47.6 & 477.2/573.0 \\
\rownumber & \texttt{0658600201} & 0.078 & 38.3/39.7 & 478.3/572.3 \\
\rownumber & \texttt{0674600601} & 0.002 & 9.0/9.4 & 483.2/573.8 \\
\rownumber & \texttt{0674600701} & 0.003 & 12.8/13.5 & 484.9/575.0 \\
\rownumber & \texttt{0674600801} & 0.003 & 17.9/18.2 & 481.4/574.1 \\
\rownumber & \texttt{0674601001} & 0.003 & 20.0/21.5 & 480.9/573.7 \\
\rownumber & \texttt{0674601101} & 0.003 & 10.1/10.7 & 480.4/573.8 \\
\hline
\end{tabular}
\caption{Properties of the XMM observations of the Galactic Center used in our analysis. 
We have only used observations with centers located
within 0.5' around Sgr A*. The difference in FoVs between MOS1 and MOS2 cameras
is due to the loss CCD6 in MOS1 camera, see~\cite{Abbey:06,MOS1-CCD6} for details.
  \newline
  $^a$ Observation discarded from our analysis due to flares in Sgr a*, see Fig.~\protect\ref{fig:gc-rate-2-6}
  and~\cite{Porquet:08}.}
\label{tab:GC-observations} 
\end{table*}

\begin{figure}[tp!]
  \centering
  \includegraphics[width=.75\linewidth]{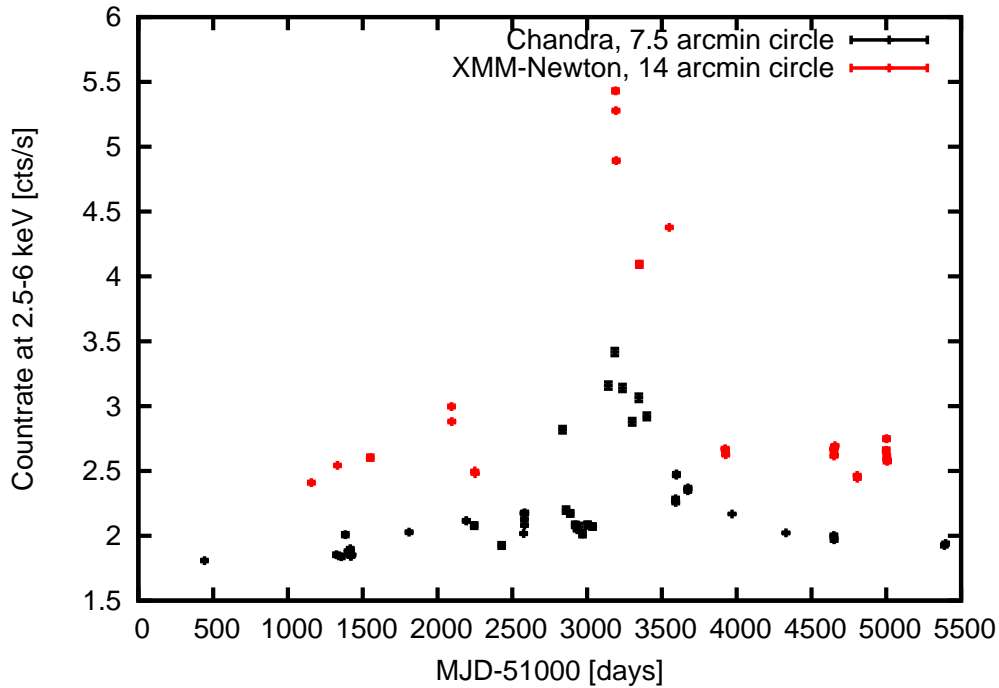}
  \caption{Average count rates on regions centered in Sgr a* using
    \textit{XMM-Newton} (red) and \textit{Chandra} (black).  The enhancement
    at MJD 54000-54500 are due to strong flaring activity of Sgr a*,
    see~\cite{Porquet:08} for details. 5 \textit{XMM-Newton}
    observations during this flaring period were discarded from our analysis,
    see Table~\protect\ref{tab:GC-observations} for details.}
  \label{fig:gc-rate-2-6}
\end{figure}

\begin{table*}[t!]
\footnotesize
\centering
  \begin{tabular}[c]{lcccccc}
    \hline
    Ion & Position & Upper level & Lover level & Emissivity & T$_e$ peak & Relative intensity\\
        &  keV     &             &             & ph cm$^{3}$ s$^{-1}$  & keV & \\
\hline
Ca XIX	 & 3.902 &  7 & 1 & 3.913e-18 & 2.725e+0 & 0.59 \\
Ca XIX	 & 3.883 &  5 & 1 & 6.730e-19 & 2.725e+0 & 0.10 \\
Ca XIX	 & 3.861 &  2 & 1 & 1.242e-18 & 2.165e+0 & 0.19 \\
\hline
Ar XVII	 & 3.685 & 13 & 1 & 8.894e-19 & 1.719e+0 & 0.13 \\
Ar XVII	 & 3.683 & 11 & 1 & 3.729e-20 & 1.719e+0 & 0.01 \\
Ar XVII	 & 3.618 & 10077 & 2 & 3.627e-20 & 1.366e+0 & 0.01 \\
Ar XVII	 & 3.617 & 10078 & 3 & 9.355e-20 & 1.366e+0 & 0.01 \\
\hline
Ar XVIII & 3.323 &  4 & 1 & 4.052e-18 & 3.431e+0 & 0.61 \\
Ar XVIII & 3.318 &  3 & 1 & 2.061e-18 & 3.431e+0 & 0.31 \\
S XVI	 & 3.276 & 12 & 1 & 9.146e-19 & 2.165e+0 & 0.14 \\
\hline
Ar XVII	 & 3.140 &  7 & 1 & 6.604e-18 & 1.719e+0 & 1.00 \\
Ar XVII	 & 3.126 &  6 & 1 & 7.344e-19 & 1.719e+0 & 0.11 \\
Ar XVII	 & 3.124 &  5 & 1 & 1.018e-18 & 1.719e+0 & 0.15 \\
S XVI	 & 3.107 &  7 & 1 & 3.126e-18 & 2.165e+0 & 0.47 \\
S XVI	 & 3.106 &  6 & 1 & 1.584e-18 & 2.165e+0 & 0.24 \\
Ar XVII	 & 3.104 &  2 & 1 & 2.575e-18 & 1.719e+0 & 0.39 \\
S XV	 & 3.101 & 37 & 1 & 7.252e-19 & 1.366e+0 & 0.11 \\
S XV	 & 3.033 & 23 &	1 & 1.556e-18 & 1.366e+0 & 0.24 \\
\hline
   \end{tabular}
  \caption{\label{tab:lines} List of astrophysical lines at 3-4 keV expected in our model. Basic line 
  parameters such as energy, type of ion, type of transition -- are taken from AtomDB database.
  Only the strongest lines
  are shown.
  Close lines of the same ion are grouped with horizontal lines.
} 
\end{table*}

\section{Dark Matter Profiles of the Milky Way}
\label{sec:profiles-milky-way}

The distribution of dark matter in galaxies, galaxy groups and galaxy clusters
can be described by several density profiles.  In this work we concentrated on
four popular choices for dark matter density profiles.

\begin{asparaenum}[\bf I.]
\item 
Numerical (N-body) simulations of the cold dark matter model have shown that
the dark matter distribution in all relaxed halos can be fitted with the
universal Navarro-Frenk-White (NFW) profile~\cite{Navarro:96}
\begin{equation}
  \rho_\nfw(r) = \frac{\rho_s r_s}{r(1+r/r_s)^2}\label{eq:rho_NFW}
\end{equation}
parametrised by $\rho_s$ and $r_s$. 

\item 
The Burkert (BURK) profile~\cite{Burkert:95} has been shown to be successful in
explaining the kinematics of disk systems (e.g.~\cite{Gentile:04}):
\begin{equation}
\label{eq:rho_BURK}
\rho_{\textsc{burk}}(r) = \frac{\rho_Br_B^3}{(r_B+r)(r_B^2+r^2)}.
\end{equation}
\item 
Another common parametrizations of cored profiles are given by the
pseudo-isothermal (ISO) profile~\cite{Kent:86}
\begin{equation}
\rho_\iso(r) = \frac{\rho_c}{1 + r^2/r_c^2}.
  \label{eq:rho_ISO}
\end{equation}

\item
The profile found by~\cite{Moore:99} from simulations is described by:
\begin{equation}
\label{eq:rho_moore}
\rho_{\textsc{moore}}(r)=\frac{\rho_c}{\sqrt{r/r_s}(1+\sqrt{r/r_s})}
\end{equation}

\item
\cite{Binney:01} found a profile from lensing data of the MW with the following general shape 
(BE in the following):
\begin{equation}
\label{eq:rho_BE}
\rho_{\textsc{BE}}(r)=\frac{\rho_c}{(r/r_s)(1+(r/r_s))^{2.7}}
\end{equation}

\end{asparaenum}

Because we reside in the inner part of Milky Way dark matter halo, it is the
only object whose dark matter decay signal would be spread across the whole
sky.  The dark matter column density for the Milky Way halo can be calculated
using the expression~\cite{Boyarsky:06d} \eq{\S_\dm^{MW}(\phi) =
  \int\limits_0^\infty \rho_\dm\left(r(z,\phi)\right)dz\label{eq:colden_Galaxy}} 
  where $r(z,\phi)=\sqrt{r_\odot^2+z^2-2 z r_\odot\cos\phi}$ is the distance from the galactic center 
  with $z$ the distance along the line of sight and $\phi$ the angle away from the GC for an observer 
  at earth (itself at $r_\odot$ from the GC). Expressed in galactic
coordinates $(l,b)$ 
\begin{equation}
\cos\phi = \cos b \cos l.
\label{eq:36}
\end{equation}
 It can be seen (e.g.~\cite{Boyarsky:06c,Boyarsky:07b,Boyarsky:06d}) that the
function $S_\dm^{MW}$ can change only by a factor of few, when moving from the
Galactic center ($\phi = 0^\circ$) to the anti-center ($\phi =
180^\circ$). That is, the Milky Way contribution to the decay is an all-sky
signal.

The flux received at earth produced by dark matter decaying inside the cone of view, we can approximate by
\begin{equation}
F_\dm^{FoV} = \S_\dm^{MW}(\phi) \Omega \Gamma / 4\pi
\end{equation}
in photons s$^{-1}$ cm$^{-2}$, with $\Omega$ the size of the field of view in $sr$, $\Gamma$ the decay 
width and the $4\pi$ to complete the distance modulus (the distance is already included in the $\Omega$). 

The exact solution, taking into account the varying density over the field of view, is
\begin{align}
F_\dm^{FoV} &= \Sigma_\dm^{FoV} \Gamma / 4\pi \\
\Sigma_\dm^{FoV} &= 2\pi \int\limits_{\phi=0}^{\phi=\omega} \int\limits_{z=o}^{z=\infty} 
\frac{\rho(r(z,\phi))}{z^2} z^2 \sin(\phi) d\phi dz
\label{eq:full-mproj}
\end{align}
for a circular field of view centered on the GC, with a radius of $\omega$.

The mass modeling of the Milky Way is continuously updated and improved~(see
e.g.~\cite{Nesti:13, Deason:12, Bernal:11, McMillan:11, Sofue:08, Xue:08, Smith:06, Battaglia:06, 
Alcock:95, Merrifield:92, Weber:09}).
In Table~\ref{tab:MW-profiles-new} we summarize recent results. 
We are interested in predicting the flux from dark matter decay based on the dark matter content. 
Therefore, using the DM distributions in the MW as reported in this table, we compute $\Sigma_\dm^{FoV}$ 
for the galactic center and blank sky observations. In the galactic center case, we perform the integral 
in eq.~\ref{eq:full-mproj} for $\omega=14'$, and then correct the results for detector gaps with the ratio 
of the exposure-weighted average FoV size (corrected for detector gaps) to the size of an ideal 14' FoV. 
For the blank sky dataset, we computed $\S_\dm^{MW}\Omega$ (see eq.~\ref{eq:colden_Galaxy}) for each blank 
sky pointing (each with its own $\phi$), therefore assuming that so far away from the GC the DM density does 
not vary appreciably over the FoV, and take the exposure and FoV weighted average of all those pointings. 
It is then, just like the case for the GC, corrected for detector gaps.

Regarding the mass modeling of the Galactic Center, there are additional complications. Firstly, 
even tough according to~\cite{Donato:09, Gentile:09}, the central surface densities of spiral 
galaxies are comparable, our field-of-view is only 14' in radius which translates to a physical 
scale of order 30~pc at the center of the halo, which is much smaller than one scale length. 
It is unfortunately not possible to observationally determine the DM distribution of the Milky Way 
within about 3~kpc from the halo center. Secondly, at these small scales, baryons dominate the mass 
budget and baryon physics may play an important role in shaping the DM distribution, in addition to 
possible warm dark matter effects. However, the extent of the influence of the processes is not well known. 
Thirdly, the central 3 kpc of the NFW distributions in Table~\ref{tab:MW-profiles-new} contribute 
between roughly 80\% (least concentrated) to 90\% (most concentrated) of the total $\Sigma^{FoV}_{DM}$ 
for the GC observations. Therefore the best we can do is extrapolate profiles measured at larger radii 
down to the lower radii. We remain agnostic about the very central DM distribution and assume that 
uncertainty is enclosed within the spread in the different types of profiles that we already examined.

Recently, \cite{Lovell:14} analysed the high-resolution Aquarius simulations specifically in 
order to predict dark matter decay fluxes. Milky Way and Andromeda-like halos from these 
simulations were selected, and the fluxes determined based on the exposure times and position 
angles as used in this work and in~\cite{Boyarsky:14a}. Since the flux in this case is determined 
solely from the mass inside the field-of-view and the assumed DM particle lifetime, 
flux and projected mass are interchangeble in this study. This produced a range of fluxes that are 
in agreement with our projected mass brackets for the GC, and the flux ratios of the GC to M31, and 
GC to blank-sky. The confidence ranges from~\cite{Lovell:14} are tighter than our literature-brackets, 
therefore we retain the latter in all joint analyses. 

To round of this discussion about the dark matter masses, we shortly touch upon the dark matter content 
of Perseus and Andromeda in order to compare our observations in Figure 2 of our paper.
As for the Milky Way, we compile available literature profiles of these objects and use those to determine 
the total dark matter mass present in the field of view of our observations~\cite{Boyarsky:14a}. 
This is a more straightforward calculation as the physical size of these objects is much smaller than 
their distance to us. We compute the enclosed projected mass of these literature profiles within the 
field of view (corrected for detector gaps), weighting by the exposure time of the different exposures, 
and then divide by the distance to the object squared to arrive at $\Sigma_\dm^{Perseus}$ and $\Sigma_\dm^{M31}$. For Perseus, we consider the profiles as determined by~\cite{Reiprich:02, Chen:07, Simionescu:12, Storm:2012ty, Ettori:02, Wojtak:07}, and those by~\cite{Klypin:01, Geehan:05, Widrow:05, Seigar:06, Tempel:07, Chemin:09, Corbelli:09} for Andromeda.

\begin{table*} [!tb]
  \centering 
\begin{tabular}[c]{|l|c|c|c|c|c|c|c|c|c|c|c|c|c|c|c|c|c|c|c|}
  \hline
  Authors & Profile & $r_\odot$ & $\rho_\ast$ & $r_\ast$ & $\Sigma^{FoV}_{DM,GC}$ & $\Sigma^{FoV}_{DM,BS}$ & GC/BS ratio \\
  & & kpc & $10^6 M_\odot/kpc^3$ & kpc & $10^{-3} M_\odot / pc^2$ & $10^{-3} M_\odot / pc^2$ & \\
  \hline
  \hline
Smith et al. 2007~\cite{Smith:06}$^a$ &  NFW  & 8 & 25.2$^{+6.2}_{-3.8}$ & 10.6$^{+0.8}_{-0.6}$ & 142.6$^{+38.2}_{-20.9}$ & 5.6$^{+1.3}_{-0.7}$ & 25.6$^{+11.6}_{-7.8}$ \\
 &  NFW  & 8 & 1.4$^{+1.2}_{-0.5}$ & 39.6$^{+4.5}_{-3.2}$ & 35.2$^{+11.6}_{-5.7}$ & 3.5$^{+1.0}_{-0.5}$ & 10.0$^{+5.5}_{-3.5}$ \\
 \hline
Weber \& de Boer 2010~\cite{Weber:09} &  NFW$^b$  & 8.33 & 20.4$^{+17.11}_{-6.4}$ & 10.8$^{+3.4}_{-3.4}$ & 118.0$^{+30.8}_{-15.8}$ & 4.5$^{+0.4}_{-0.4}$ & 26.2$^{+10.3}_{-5.5}$ \\
 &            NFW$^b$ & 8.33 & 6.32$^{+1.26}_{-0.78}$ & 25.2$^{+4.6}_{-4.6}$ & 95.1$^{+10.1}_{-8.4}$ & 7.1$^{+0.6}_{-0.6}$ & 13.4$^{+2.8}_{-2.1}$ \\
 &  BE  & 8.33 & 6.58$^{+1.3}_{-1.3}$ & 10.2 & 22.0$^{+3.5}_{-4.4}$ & 4.0$^{+0.8}_{-0.8}$ & 5.5$^{+2.4}_{-1.8}$ \\
 &  Moore$^c$  & 8.33 & 6.58$^{+1.3}_{-1.3}$ & 30 & 306.3$^{+60.7}_{-60.7}$ & 3.9$^{+0.8}_{-0.8}$ & 77.8$^{+38.5}_{-25.8}$ \\
 &  PISO$^d$  & 8.33 & 5.264$^{+1.04}_{-1.04}$ & 5 & 11.4$^{+2.6}_{-1.8}$ & 3.5$^{+0.7}_{-0.7}$ & 3.2$^{+1.7}_{-0.9}$ \\
 \hline
Battaglia et al. 2005, 2006~\cite{Battaglia:05,Battaglia:06}$^e$ &  NFW  & 8 & 11.4 & 14.86$^{+0.71}_{-0.49}$ & 95.1$^{+5.3}_{-3.5}$ & 5.0$^{+0.5}_{-0.3}$ & 19.0$^{+2.4}_{-2.3}$ \\
 &  NFW  & 8 & 11.4 & 16.12$^{+0.44}_{-0.46}$ & 103.9$^{+3.5}_{-2.6}$ & 5.8$^{+0.3}_{-0.3}$ & 17.8$^{+1.7}_{-1.3}$ \\
 \hline
McMillan 2011~\cite{McMillan:11} &  NFW  & 8.29 & 8.49$^{+2.85}_{-1.59}$ & 20.2$^{+4.3}_{-4.3}$ & 99.5$^{+11.9}_{-8.4}$ & 6.4$^{+0.5}_{-0.5}$ & 15.5$^{+3.2}_{-2.3}$ \\
\hline
Nesti \& Salucci 2013~\cite{Nesti:13} &  NFW  & 8.08$\pm$0.2 & 13.8$^{+20.7}_{-6.6}$ & 16.1$^{+12.2}_{-5.6}$ & 125.9$^{+75.6}_{-26.1}$ & 7.0$^{+3.5}_{-1.3}$ & 18.0$^{+17.3}_{-8.5}$ \\
 &  BURK  & 7.94$\pm$0.3 & 4.13$^{+4.4}_{-1.1}$ & 9.26$^{+4.0}_{-3.0}$ & 22.9$^{+21.4}_{-5.0}$ & 7.3$^{+11.5}_{-2.1}$ & 3.2$^{+5.5}_{-2.2}$ \\
 \hline
Xue et al. 2008~\cite{Xue:08}$^f$ &  NFW$^g$  & 8 & 4.2$^{+0.3}_{-0.3}$ & 21.9$^{+1.1}_{-1.4}$ & 53.7$^{+7.9}_{-6.2}$ & 3.8$^{+0.7}_{-0.6}$ & 14.2$^{+4.9}_{-3.6}$ \\
 &  NFW$^g$  & 8 & 4.4$^{+0.2}_{-0.4}$ & 20.8$^{+1.1}_{-1.0}$ & 52.8$^{+7.0}_{-6.2}$ & 3.6$^{+0.6}_{-0.6}$ & 14.7$^{+5.1}_{-3.6}$ \\
 &  NFW  & 8 & 0.99$^{+0.76}_{-0.45}$ & 41.1$^{+6.4}_{-5.8}$ & 25.5$^{+12.3}_{-7.9}$ & 2.6$^{+1.0}_{-0.6}$ & 9.8$^{+9.4}_{-4.9}$ \\
 &  NFW  & 8 & 0.47$^{+0.32}_{-0.18}$ & 60.2$^{+7.2}_{-7.2}$ & 18.5$^{+8.8}_{-5.3}$ & 2.3$^{+0.8}_{-0.6}$ & 8.2$^{+8.0}_{-3.9}$ \\

  \hline
\end{tabular} 
\caption{Overview of dark matter distributions as determined in the literature. $r_\ast$ and $\rho_\ast$ 
refer to the relevant characteristic radius and density for that particular type of profile. 
Where the profile was given in a different parametrization of the same profile 
(for example, concentration and virial mass), the values have been converted to $r_\ast$ and $\rho_\ast$. 
The errors given are 1$\sigma$, which are naively converted from the error range given in that work if 
that range was not 1$\sigma$. $\Sigma^{FoV}_{DM,GC}$ (see Eq.~\ref{eq:full-mproj}) is the integral 
over the density of the galactic center inside the field of view of our observations, divided by the 
distance squared to each infinitesimal mass. $\Sigma^{FoV}_{DM,BS}$ is the same, but for the blank-sky 
dataset from Boyarsky et al. 2014. The errors on these projected mass densities are either $0.5\sigma$ 
to account for the degeneracy between the 2 parameters of the DM distribution, or $1\sigma$ if the fit 
from that study fixed one of those parameters (for example using a scaling relation between $c$ 
and $M_{vir}$). \\
\textit{a)} the two descriptions are using different baryonic disks. \textit{b)} some baryonic parameters 
are fixed in the fits. These two NFW's are the two extremes with reasonably good fits. \textit{c)} the 
Moore model is very cuspy by design. \textit{d)} the pseudo-isothermal sphere has an almost flat 
profile in the center. \textit{e)} the second NFW takes anisotropy into account, and is a better fit 
that the first. \textit{f)} analysis calibrated on two different simulations. \textit{g)} includes 
an adiabatic correction. 
}
\label{tab:MW-profiles-new} 
\end{table*}

\section{Ion Abundances and Emission Lines}
\label{sec:abundances}

The Galactic Center is an object with a complicated signature in the X-rays. As \cite{Muno:04b} show, 
not only does the GC show multi-temperature components in the X-ray spectra, 
these components also vary quite dramatically spatially over the field-of-view of Chandra, 
which is about half as large as that of XMM-Newton. The low temperature component as measured by 
\cite{Muno:04b} typically has values of $0.7$ -- $0.9$~keV, while the high temperature component can 
be as hot as $6$ -- $9$~keV. The spatial variations in the elemental abundances of Si, S, Ar and Ca 
are reported to be as high as a factor 2 or 3, with only Fe having a reasonable homogeneous distribution. 
Our integrated spectrum of the entire inner 14' of the GC therefore will be a superposition of all 
these components, complicating our analysis significantly. 

Restricting our modelling to the cleaner parts of the spectrum, 2.8--6.0 keV, we could find a reasonable 
fit using a single-temperature \tt vvapec \normalfont component with the elemental lines added manually 
as gaussians, and a folded powerlaw to account for non-thermal emission. No satisfactory two-temperature 
fits were found for temperatures in the range given by \cite{Muno:04b}, even when extending the energy range 
of our analysis\footnote{The two-temperature fit can be made satisfactory e.g. by adding 1.2\% 
systematic error in quadrature -- a value much larger than the typical systematic errors for 
line-like uncertainties ($\sim $0.5\%, see Sec.~5 of~\cite{Iakubovskyi:13} for details). When adding such 
large errors, the \tt vvapec \normalfont temperatures become consistent with previous works 
(e.g.~\cite{Muno:04b}) and the abundances of S, Ar, Ca and K are 0.8-1.2, 1.2-1.8, 1.6-2.4 and 0.3-3.4~Solar 
values at 68\% level, respectively, in full accordance with~\cite{Jeltema:14a}.}.
We did not consider more than two temperature components, because it introduces too many degeneracies. 

As mentioned, the emission lines from heavy ions are added by hand. We start with the strongest lines known 
(see Table~\ref{tab:lines}), and work our way down so long as the fit requires it. As mentioned, the line 
detected 
at 3.539~keV might be influenced by the Ar~XVII complex at 3.685~keV and the K~XVIII lines at 3.515 and 
3.47~keV. To explain the 3.539~keV line with Ar~XVII, the width of this line should be much larger 
(95 -- 130~eV) than what can be expected from the instrumental response based on simulations of this Ar 
complex. In addition, the flux in this Ar~XVII complex would be higher than that of the same ion at 3.13~keV, 
which should not be possible based on the atomic data in Table~\ref{tab:lines}. 

For the K~XVIII lines, it is unfortunately not possible to constrain their
contribution to the 3.539~keV line in the same way as for Ar~XVII, since we do
not have other, stronger, detected lines of the same ion in our spectrum. In
this case, one may attempt to predict the ratio of K~XVIII flux to the fluxes
of ions of other elements such as Ar~XVII, Ca~XIX, Ca~XX, S~XVI, etc. based on
temperature and relative abundances. Since the GC emission consists of many
different temperature and abundance components, it should be necessary to
compute an estimate of the K~XVIII flux for many different combinations of
temperature and abundance. Based on the flux of each of the different detected
strong lines in turn and assuming solar abundance (similarly to the analysis
of Section~4 of~\cite{Bulbul:14}), the predictions for the K~XVII flux can
vary by more than an order of magnitude. Even without considering deviations
from solar abundance (which may be as large as a factor 3~\cite{Jeltema:14}),
the detected flux in the 3.539~keV line falls within these predictions for a
respectable fraction of these physically plausible scenarios.  It is therefore
not possible \emph{based on the GC data alone} to exclude the astrophysical
origin of this 3.539~keV line in the GC.

\end{document}